\documentclass[preprintnumbers,superscriptaddress]{revtex4}
\usepackage{epsf}
\usepackage{graphicx}
\usepackage{dcolumn}
\usepackage{bm}

\def\be{\begin{equation}}
\def\ee{\end{equation}}
\def\ba{\begin{eqnarray}}
\def\ea{\end{eqnarray}}

\begin{document}

\draft

\title{Mesoscopic oscillations of the conductance of
 disordered metallic samples as a function of temperature}

\author{B. Spivak}

\affiliation{Physics Department, University of Washington,
Seattle, WA 98195, USA}

\author{A. Zyuzin}

\affiliation{A.F.Ioffe Institute, 194021 St.Petersburg, Russia}

\author{D.H. Cobden}
\affiliation{Physics Department, University of Washington, Seattle, WA 98195, USA}

\begin{abstract}

We show theoretically and experimentally that the conductance of small disordered samples exhibits random
oscillations as a function of temperature. The amplitude of the oscillations
 decays as a power law 
of temperature, and their
characteristic period is of the order of the temperature
itself.
\end{abstract}

\maketitle

\draft

At low temperatures the conductance of small disordered metallic samples
fluctuates from sample to sample. There are two contributions to the
amplitude of the fluctuations. The first is associated with a classical
effect: the Drude conductivity depends on the concentration of impurities,
which fluctuates in space. The second effect is due to electron quantum
interference. As a consequence of the latter, the conductance
 of an individual sample exhibits random oscillations as a function of
  external magnetic field and chemical potential.
The goal of this communication is to point out that the conductance $G(T)$
 of an individual mesoscopic metallic sample also oscillates as a function of {\it temperature}. 
 
 The well known picture of mesoscopic fluctuations of the conductance between samples, and
of oscillations of the conductance of individual samples as a function of magnetic field and chemical potential, is as follows. When
the sample conductance $G\gg e^{2}/\hbar $ is large and at zero temperature 
$T=0$, the variance of the interference contribution is universal, 
\begin{equation}
\langle (\delta G)^{2}\rangle =\alpha \frac{e^{4}}{\hbar ^{2}},
\end{equation}
and independent of sample size \cite{LeeStone,Altshuler}. Here $\delta
G=G-\langle G\rangle $, the 
brackets $\langle \rangle$
denote averaging over a random scattering potential, and $\alpha $
 is a coefficient of order unity which depends on the
dimensionality of the sample and its geometry. One can get Eq. 1 by calculating the diagram shown in
Fig.1. (We use a standard diagram technique for averaging over random
scattering potential \cite{abricosov}.) The conductance of an individual
sample, $G({\bf H})$, exhibits random sample specific oscillations as a
function of external magnetic field ${\bf H}$ \cite{Web,LeeStone}. We
will consider for example the sample geometry  shown in the inset of
Fig.2, and assume that the sample size $L$ is much larger than the elastic mean
free path, $L\gg l$. If the magnetic length $L_{H}\ll L$ and at $T=0$, the
amplitude of the oscillations is  given by Eq. 1, while their
characteristic period is  $H^{\ast }\sim \Phi _{0}/L^{2}$, where
 $\Phi _{0}=h /ec$ is the flux quantum. This
statement follows from the magnetic field dependence of the correlation
function, 
\begin{equation}
\langle (\delta G(\mathbf{H}+\Delta \mathbf{H})\delta G(\mathbf{H}))\rangle
\sim \frac{e^{4}}{\hbar ^{2}}\Gamma (\Delta \mathbf{H}).
\end{equation}
At $\Delta H\gg H^{\ast }$ the correlation function has the asymptotic behavior
 $\Gamma (\Delta H)\sim L_{\Delta H}/L$
 and  approaches zero. This  can be
shown by calculating the diagram in Fig.1, assuming that the
inner solid lines correspond to electron Green functions at magnetic field ${\bf H}$
while the outer solid lines correspond to Green functions at  ${\bf H}+\Delta {\bf H}$.
 The oscillations of
the conductance as a function of $H$ in the regime where $L_{T}\gg L_{H}\gg
L $ were discussed in \cite{ZyuzinSpivak}. Here $L_{T}=\sqrt{D/T}$, where
$D$ is the diffusion coefficient of the metal.
 For example in the three-dimensional (3d) case the amplitude
of the oscillations decays as $L^{-1}_{H}$ while their
period  is of order $H$. Thus in this
regime the typical period of the oscillations decreases while  the
derivative $dG/dH$ diverges as $H\rightarrow 0$. To get these results one
has to assume that the electron diffusion coefficient in the leads is the
same as in the sample.

The oscillations mentioned above are of a single-particle interference
nature. Contributions to $\delta G$ from electron wave functions with
different energies, generally speaking, have different signs. 
As the temperature $T$ increases, cancellation of contributions  at different energies
 becomes more effective,
leading to a decay of the amplitude of the mesoscopic oscillations.

In this article we show that the temperature
dependence of the conductance $G(T)$ of an individual sample is actually a
non-monotonic function of the temperature $T$ and exhibits random sample
specific oscillations. The characteristic period of the oscillations $
T^{*}$ is of the order of the temperature itself, that is, 
\begin{equation}
T^{*} \sim T.
\end{equation}

 To prove the existence of the oscillations we calculate the correlation
function 
\begin{equation}
\langle \delta G(T_{1})\delta G(T_{2})\rangle =\left( \frac{2e^{2}D}{\pi h
}\right) ^{2}\left( \frac{1}{L}\right) ^{4-d}\int_{0}^{\infty} d {\bf q}
dt\left[ \frac{2}{Dq^{2}+\tau _{\phi }^{-1}}+t
\right] \exp [-(Dq^{2}+\tau _{\phi }^{-1})t]B(tT_{1})B(tT_{2}),
\end{equation}
where $\tau _{\phi }$ is the
electron phase breaking time, and 
\begin{equation}
B(z)=\frac{\pi z}{\sinh \pi z}.
\end{equation}%

It follows from Eq. 4 that in the limit $T\gg T^{\prime }\gg D/L^{2}$ 
\begin{equation}
\langle {\delta G(T))^{2}}\rangle =\alpha _{1}\frac{e^{4}}{\hbar ^{2}}
\left\{ 
\begin{array}{ll}
\frac{D^{1/2}}{T^{1/2}L} & \text{ for $d=3$} \\ 
\frac{D}{TL^{2}}\ln (\tau _{0}T) & \text{for $d=2$}
\end{array}
\right.
\end{equation}
and 
\begin{equation}
\langle \delta G(T^{\prime })\delta G(T)\rangle =\alpha _{1}\frac{e^{4}}{
\hbar ^{2}}\left\{ 
\begin{array}{ll}
\frac{D^{1/2}}{T^{1/2}L}(1-0.4(\frac{T^{\prime }}{T})^{2}) & \text{ for $d=3$} \\ 
\frac{D}{T L^{2}}\ln (\tau _{0}T)(1-0.8(\frac{T}{T^{\prime }})^{2}) & 
\text{for $d=2$}
\end{array}
\right.
\end{equation}
 Here 
 $\tau _{0}=\min \{\tau _{\phi},D/L^{2}\}$, and $\alpha _{1}$ is a coefficient of order unity which depends on the sample geometry. To get Eq. 6 one can calculate
the diagram shown in Fig. 1 where the electron Green functions in the inner
and the outer loops are taken at the same temperature $T$. To get Eq. 7 one has
to take the Green functions in the inner loop at $T^{\prime}$ and in the
outer loop at $T$.

The existence of the oscillations of the conductance $G(T)$ as a function of $T$,
and the fact that $\delta G(T)$  changes  sign, follow from the facts that the
 powers of temperature in the
denominators of Eq. 6 and Eq. 7 are the same, and that Eq. 7 is 
independent of $T^{\prime }$ for $T\gg T^{\prime }$. (For example a typical monotonic form $\delta G(T)\sim
AT^{-\gamma }$ cannot satisfy both Eq. 6 and Eq. 7 even if the coefficient $A$ 
has a random sample-specific sign.) A typical realization of the
temperature dependence of the conductance has the form 
\begin{equation}
\delta G(T)=\langle (\delta G(T))^{2}\rangle ^{1/2}F(T),
\end{equation}%
where the function $F(T)$ randomly oscillates about zero with a
characteristic period of order $T$. 

In the opposite limit $T\ll D/L^{2}$, (ie. $L_{T}\gg L$), the temperature dependence of
 $\delta G(T)$ depends
on the properties of the leads. In the limit
 $D_{L}/ D\rightarrow \infty$, where $D_{L}$ is the diffusion
coefficient in the leads,  the function $\delta G(T)$ vanishes monotonically 
 as $T\rightarrow 0$.
However, if $D_{L}\sim D$ the $T$-dependence of  $G(T)$ has similar features to its $H$-dependence in the limit $
L_{T}\gg L_{H}\gg L$ \cite{ZyuzinSpivak}. That is, $\delta G(T)$ exhibits random
oscillations whose amplitude decays as $T\rightarrow 0$. The period of
the oscillations for $T\ll D/L^{2}$ is again of order $ T^{*}\sim T$. The
latter statement follows from the $T$ dependence of the following
correlation functions at $T^{\prime }\ll T $, 
\begin{equation}
\langle (\frac{d\delta G(T)}{dT})^{2}\rangle \sim \frac{1}{T^{3-d/2}}
\end{equation} 
and
\begin{equation}
\langle \frac{d\delta G(T^{\prime })}{dT^{\prime }}\frac{d\delta G(T)}{dT}
\rangle \sim -\frac{T^{\prime }}{T^{4-d/2}}       
\end{equation}
 which can be obtained by calculation of diagrams in Fig.1 as in
\cite{ZyuzinSpivak}. According to Eqs. 9 and 10 the value of the derivative 
$dG/dT $ diverges as $T\rightarrow 0$. This is correct as long as $L_{\phi
}\gg L_{T}$, where $L_{\phi}=\sqrt{D\tau_{\phi}}$ is the electron phase breaking length. 
The latter inequality holds if the value of $L_{\phi }$ is
determined by electron-electron or electron-phonon scattering \cite{Aronov}.
The qualitative temperature dependence of $\delta G(T)$ in this case is shown in Fig.2.

At very low temperatures the value of $L_{\phi }$ is determined by the
paramagnetic impurities in the sample and is temperature independent as long
as the Kondo effect and exchange between paramagnetic spins are not
significant. In the case $L_{\phi }\ll L_{T}$ the amplitude of the
oscillations of $dG/dT$ decays as $T\rightarrow 0$. Thus the typical
amplitude of the oscillations of the derivative $dG/dT$ has a maximum when 
$L_{T}\sim L_{\phi }$; and the total number of  oscillations is of order 
$\ln (T\tau _{s})$, where $\tau _{s}$ is the spin relaxation time.

The oscillations as a function of temperature discussed above should be present in 
any thermodynamic or transport property of mesoscopic metallic samples.

To test the theory we study some measurements of conductance oscillations in
a silicon MOSFET as a joint function of gate voltage $V_g$ and temperature $
T $. The chosen device has a square channel of length and width $L \sim 1 \mu
$m. The oxide thickness is 25 nm, giving a gate capacitance per unit area of 
$8.6 \times 10^{11} e$ cm$^{-2}$ V$^{-1}$. The source and drain contacts are
n++ doped silicon. The average conductance, measured by passing
an ac current of 5 nA, varies approximately linearly from $19e^2/h$ at 
$V_g = 4$ V to $26e^2/h$ at $V_g = 5$ V.  For practical purposes, the device
behaves as a square of disordered 2D electron gas with mobility $\mu \sim
2000$ cm$^2$ V$^{-1}$ s$^{-1}$ and momentum scattering length $l \sim 30$ nm,
having 3D metallic contacts. Measurements of conductance oscillations as a
function of magnetic field \cite{Cobden} indicate that the channel is phase
coherent at the base temperature of 35 mK achieved on the dilution
refrigerator.

The data presented in Fig. 3 are derived from sweeps of $V_g$ at a series of
temperatures between 35 mK and 1.2 K. (Note that a
constant perpendicular magnetic field of 0.1 T was present in all
measurements.) A smooth monotonic background variation of the mean
conductance $\langle G\rangle$ with $V_g$ and $T$ has been subtracted 
from the raw data, so that the
quantity plotted in the figure is the deviation from this background, $%
\delta G = G - \langle G\rangle$. The sweeps show reproducible oscillations which decay
and broaden as $T$ increases, as illustrated in Fig. 3a. The variance 
$\langle(\delta G) ^2 \rangle$ and correlation gate voltage $V_{g}^{*}$ are plotted
against $T$ in Fig. 3b. Fig. 3c is a greyscale plot of $\delta G$ vs $V_g$
and $T$, where peaks are light and dips are dark. The
appearance of this plot, where individual extrema evolve steadily with $T$,
gives us confidence that the data at different temperatures can
be compared reliably.

Fig. 3d shows the variation of $\delta G$ with $\log{T}$ at a set of evenly
spaced values of $V_g$. The curves here are smooth splines passing through
the eight temperature points at each $V_g$ and extrapolating towards $\delta G =
0$ at $T \gg 1.2$ K. For clarity, the actual data points are marked as solid circles on only
one of the curves. It is apparent that $\delta G$ oscillates randomly with $T$ on a logarithmic scale, in qualitative accordance with our predictions. Over the
factor-of-30 temperature range here, at each gate voltage typically one or
two oscillations are resolved. Selected curves have been drawn in bold to
illustrate the variety of the oscillatory behavior.

It is quite surprising that these oscillations of the conductance as a
function of temperature have never been pointed out in either the
theoretical or the experimental literature.

We would like to thank D. Khmelnitskii, L. Levitov, and N. Birge for useful
discussions, C. Ford and J. Nichols for help with the experiments,
which were performed at the Cavendish Laboratory, and Y. Oowaki of
Toshiba for supplying the devices. This work was supported in part by the
National Science Foundation under Contracts No. DMR-0228014 (BS).

\newpage

\begin{figure}
\centerline{\epsfxsize=10cm \epsfbox{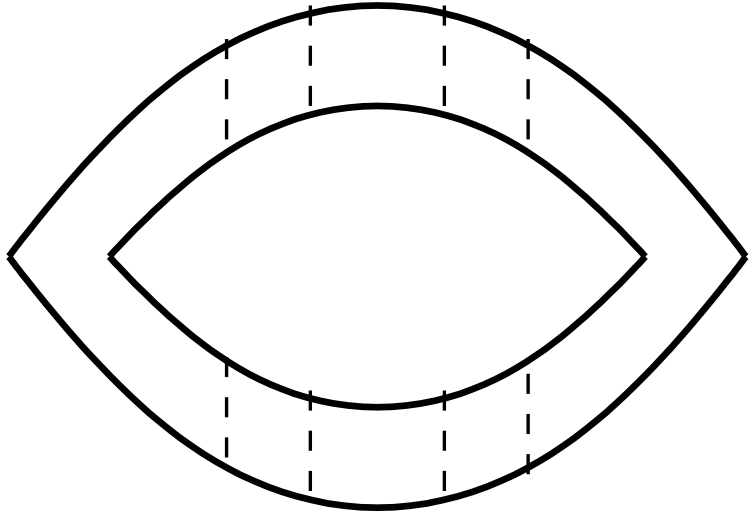}} \ 
\caption{Diagram describing the correlation function $\langle G({\bf
H},T) G({\bf H}^{\prime},T^{\prime}) \rangle$. Solid lines correspond to
electron Green functions, and thin dashed lines correspond to the correlation
function of the random scattering potential. }
\label{fig:fig1}
\end{figure}

\newpage

\begin{figure}
\centerline{\epsfxsize=10cm \epsfbox{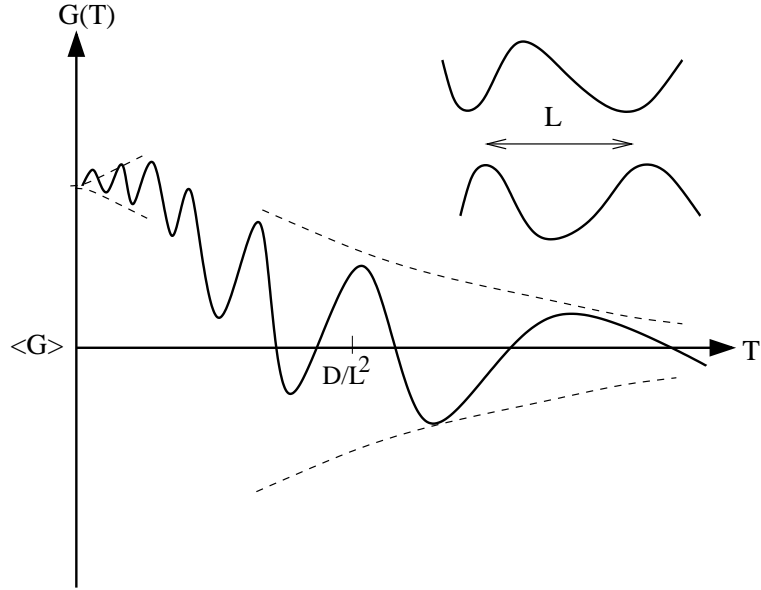}} \ 
\caption{ Typical temperature dependence of the conductance $G(T)$. 
Insert: schematic diagram of the sample. }
\label{fig:fig2}
\end{figure}

\newpage

\begin{figure}
  \centerline{\epsfxsize=10cm \epsfbox{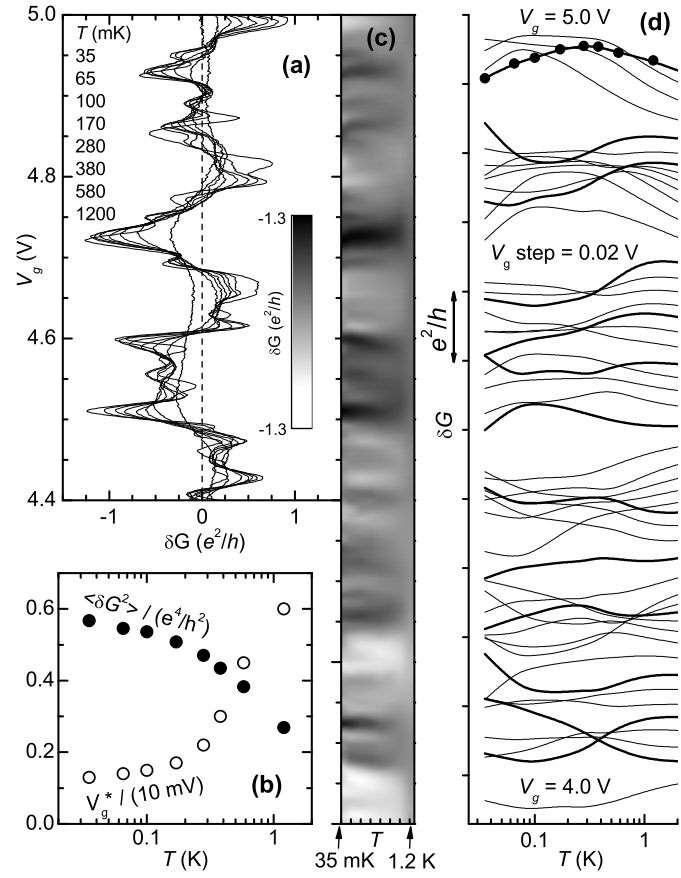}}
  \caption{Measurements of the conductance oscillations $\delta G$
in a silicon MOSFET. (a) Gate voltage ($V_g$) sweeps at a series of
temperatures ($T$) listed at the top left.  (b) Variance $\langle (\delta G)^{2}\rangle$ and correlation
gate voltage $V_{g}^{*}$ (the half-width of the autocorrelation function) of the
oscillations as a function of temperature, obtained by averaging over
$V_g$.  (c) Greyscale plot of $\delta G$ vs $T$ and $V_g$.  (d) Temperature
dependence of $\delta G$ at $V_g = 4.0, 4.02, .... 5.0$ V, with consecutive
sweeps offset vertically by $0.2 e^2/h$. } 
  \label{fig:fig3}
\end{figure}

\end{document}